\documentstyle  [12pt]{article}

\def\be{\begin{equation}}
\def\ee{\end{equation}}
\def\bea{\begin{eqnarray}}
\def\eea{\end{eqnarray}}
\def\ba{\begin{array}}

\def\ea{\end{array}}
\def\nn{\nonumber}
\begin{document}
\begin{titlepage}
\begin{flushright}
hep-th/9601084 \\
IPM-96-122\\
Jan. 1996
\end{flushright}
\begin{center}
{\bf \Large   NON-COMMUTATIVE GEOMETRY AND CHIRAL PERTURBATION
LAGRANGIAN}
\end{center}
\vskip 1cm
\centerline{ M. Alishahiha${^{a,b}}$, A. H. Fatollahi${^{a,b}}$,
K. Kaviani${^{a,c}}$\footnote {E-Mail: kaviani@vax.ipm.ac.ir}}
\vskip 1cm
\centerline {${^a}${\it Institute for Studies in Theoretical Physics and
Mathematics,}}
\centerline{\it  P.O.Box 19395-5746, Tehran, Iran }
\centerline {${^b}${\it  Department of Physics, Sharif University of
Technology,}}
\centerline{\it  P.O.Box 11365-9161, Tehran, Iran }
\centerline {${^c}${\it Department of Physics, Az-zahra University}}
\centerline{\it P.O.Box 19834, Tehran, Iran}
\vskip 2cm
\begin{abstract}
Chiral perturbation lagrangian in the framework of non-commutative geometry is
considered in full detail.  It is found that the explicit symmetry breaking
terms appear and some relations between the coupling constants of the
theory come out naturally. The WZW term also turns up on the same footing as
the other terms of the chiral lagrangian.
\end{abstract}
\end{titlepage}
\section{Introduction}

It has been shown by Connes\cite{C,CL} that the lagrangian of the Standard
Model with the Spontaneous symmetry breaking mechanism can be obtained
naturally by using the non-commutative geometrical concepts for
differentiation, integration, connection, curvature, etc.
The Higgs Field appears in non-commutative lagrangian as a generalized
gauge field in the direction of discrete dimension [1-6]. In Ref\cite{AK} it
was shown that the explicit symmetry breaking mechanism may also
be related to the existence of an additional discrete dimension.
The mass terms in the lagrangian of the Chiral Perturbation Theory (ChPL)
were obtained in the framework of Non-Commutative Geometry (NCG) up to
the 4th order of momentum.
There are two remarkable points in this approach. First, the complicated
terms in the ChPL are obtained by simple assumptions and direct calculations;
second, some relations between the coupling constants of the theory come out
naturally.
As the lagrangian of Ref\cite{AK} was limited to the 4 dimensional space
in the continuous part of the geometrical space, the
WZW term was  absent. This term can be obtained if we take a 5 dimensional
manifold. In Ref\cite{AK} also, instead of the field $\chi$ which appear
in Gasser-Leutwyler lagrangian\cite{GL}, only the $m^2$ term
(m is the meson mass) appeared. In this paper, by
a suitable definition of $\chi$ in terms of the mass matrice $M$
and the mesonic field on the other layer of space, we regenerate the original
ChPL.

This paper is divided in to four sections as follows.  In section 2 we
introduce the tools which are needed in NCG. In section 3 we will
construct the lagrangian of ChPT in the framework of NCG, and in the last
section we will study the WZW term, in this framework.

\section{A Brief Review of NCG }

There is a well-known theorem due to Gelfand and Naimark that describes
how one may substitute a compact topological space ${\cal M}$ with the
algebra ${\cal C}^\infty({\cal M})$ of complex continuous functions
defined on ${\cal M}$, which has sup norm and is commutative.
The extension of this theorem is the Connes' proposal which considers
how to define a compact, non-commutative space
in terms of a unital, non-commutative $*$-algebra ${\cal A}$\cite{C}.

The starting point of Connes approach is the K-cycle $({\cal H}, D)$, where
${\cal H}$ is the Hilbert space for representing the elements
of ${\cal A}$ as the linear operators and $D$ is called the generalized
Dirac operator. The role of K- cycle
for NCG is similar to the role of differential structure in ordinary
differential geometry. Having a K-cycle one can develop a
differential algebra $\Omega({\cal A})$ for the non-commutative space
which is equivalent to
differential geometry for the manifolds. For this purpose assume $d$ to be an
abstract differential operator which acts on elements of ${\cal A}$ and
satisfies  the Leibniz rule, with $ d1=0, d^2 a=0$ where $1,a \in {\cal A}$.
Connes defines the $p$-forms in $\Omega({\cal A})$ as:
\begin{equation}
\alpha=\sum _i \,a_0^i\, da_1^i\, ... \,  da_p^i \,\,\,\,\,\,\,\,\,;
a_0^i,a_1^i,...,a_p^i \in {\cal A} ,
\end{equation}
and also the representation of $\alpha$ in ${\cal H}$ is shown by
$\pi(\alpha)$ and is defined as:
\begin{equation}
\pi(\alpha)=\sum_{i} \pi(a^{0}_i)[D,\pi(a^1_i)]...[D,\pi(a^{p}_i)].
\end{equation}

As a simple example, consider a Euclidean spin manifold ${\cal M}$.
For  such manifold, we should take ${\cal A}$ the algebra of complex valued
functions on ${\cal M}$,
\begin{equation}
{\cal A} := {\cal C}^\infty ({\cal M}),
\end{equation}
then $D$ is the ordinary Dirac operator \footnote {We ignore the $i$'s in
our calculations because we can finally absorb them in the coupling
constants of the theory. }
\begin{equation}
D\,=\, \partial \kern -0.5em /  \,=  \gamma ^\mu \partial _\mu .
\end{equation}
According to the above definition,we have:
\begin{equation}
\pi(dg)\,\,=\,\,[\,D\,,\,g\,], \,\,\,\,\,\,\forall g \in {\cal A}
\end{equation}
which gives
\begin{equation}
\pi(dg)=[\,\partial \kern -0.5em /  \,,\,g\,]\,\,=\,\,\gamma ^\mu
\,\,\partial _\mu \,g
\equiv \gamma ({\bf d}g)\,\,,
\end{equation}
where ${\bf d}g$ is the ordinary one-form, ${\bf d}g= \partial _\mu g {\bf
d}x^\mu $.
To describe a fibre bundle over the manifold ${\cal M }$, we may choose
\begin{equation}
{\cal A}={\cal C}^\infty({\cal M}) \otimes M_N({I \kern -.6em C}),
\end{equation}
where $M_N({I \kern -.6em C})$ is the algebra of $N \times N $ complex
matrices. Also one may assume the discrete dimension for geometrical spaces.
For instance in the case of a two layer space, each layer is described by
the algebra of continuous functions ${\cal C}^\infty({\cal M})$.
The proper algebra for this geometry is:
\begin{equation}
{\cal A}={\cal C}^\infty({\cal M}) \otimes ( \,\,M_N({I \kern -.6 em C}) \oplus
M_N({I \kern -.6em C})\,\,).
\end{equation}
We may take ${\cal M}$ as a $4$-dimensional Euclidean compact manifold. To
make differential algebra, the Dirac operator can be chosen as follows.
\begin{equation}
D = \left( \matrix{ \partial \kern -0.5em /  \,\, \otimes \,\,
{ 1\kern - .4em 1} &
\gamma _5 \otimes M  \cr \gamma _5 \otimes M^{\dag}  &
\partial \kern -0.5em /  \,\, \otimes \,\, { 1\kern - .4em 1} }
\right)
\end{equation}
where $M$ is an $N\times N$ matrix. A representation of any element
$g \in {\cal A}$ in ${\cal H}$ can be taken as:
\begin{equation}
\pi(g) \equiv \left (\matrix {V(x) & 0 \cr 0 & V^\prime (x) } \right
)\,\,\,\,,\,\,g\,\,\in {\cal A}
 \end{equation}
where $V(x)$ and $V^\prime (x) $ are $N \times N $ matrices and $x$ indicates
the coordinates on ${\cal M}$. This representation contains in
itself the information about the two layers of space. $V(x)$ represents
the functions on one layer and $V^\prime (x)$ those of the
other.
Now it can easily be shown that:
\begin{eqnarray}
\pi(dg) &=& [\,D\,,\pi(g)\,]
= \left ( \matrix{\partial \kern -0.5em /  V & \gamma _5 (M\,V^\prime
\,-\,V\,M) \cr
\gamma _5 (M^{\dag }\,V\,-\,V^\prime \,M^{\dag}) & \partial \kern -0.5em /
V^\prime} \right ).
\end{eqnarray}

At the end it would be useful to introduce the unitary groups of the algebra
${\cal A}$ as:

\begin{equation}
{\cal U}({\cal A})= \{g \in {\cal A} \,\,, gg^*=g^*g =1\},
\end{equation}
where $*$ indicates the involution in ${\cal A}$.

\section{ Chiral Perturbation Theory}

Gasser and Leutwyler in their method of obtaining the
ChPL which is the effective lagrangian for low energy QCD,
stated that this lagrangian should preserve all the symmetries of QCD.
Their method however was not based on using the QCD lagrangian, instead it
was completely phenomenological.
Their basic idea was  expanding the
effective lagrangian in terms of different powers of the mesonic
field momentum i.e. in terms of different powers of the field
derivative\cite{GL,L}.
Here we use a similar idea but we utilize the generalized
derivative of non-commutative geometry for the mesonic field instead of
the ordinary one. We start
from a simple minded form of the lagrangian in terms of different powers of
$L=L_\mu {\bf d}x^\mu = (U \partial _\mu U^{\dag }) {\bf d}x^\mu $, where
$U \in SU(3)_{flavour}$ is the mesonic field (actually
$U= e ^ {i \lambda ^a \pi _a \over 2}$
where $\pi$ is the pionic field). As we shall see, in our approach all
the explicit symmetry breaking terms (mass included terms) appears naturally.
More than that we obtain some extra relations between the
coupling constants of the theory.

At the beginning we take the following algebra:
\begin{equation}
{\cal A}=({\cal C}^\infty({\cal M}) \otimes M_N({I \kern -.6 em C})) \oplus
M_N({I \kern -.6em C})
\end{equation}
which describes the geometrical space containing a 4-dimensional manifold
and a point which is separated from ${\cal M}$ by the additional discrete
dimension. The Dirac operator is taken to be (See also Ref\cite{AK} ):
\begin{equation}
D = \left( \matrix{ \partial \kern -0.5em /  \,\, \otimes \,\, { 1\kern - .4em
1}
& \gamma
^5 \otimes M  \cr \gamma^5 \otimes M^{\dag}  &
0 }
\right)
\end{equation}
where $M$ and ${ 1\kern - .4em 1}$ are $3 \times 3$ unitary matrices, and

\begin{equation}
M M^{\dag} = M^{\dag} M =m^2 1 \kern - .4em 1.
\end{equation}
If $g$ is an element in ${\cal U}({\cal A})$, the unitary group of ${\cal A}$,
then we generalize the one-form $L=(U \partial _\mu U^{\dag }) {\bf d}x^\mu $
as follow:
\begin{equation}
L=gdg^{*} \,\,\,, g \in {\cal U}({\cal A}).
\end{equation}
The representation of $g$ in the Hilbert space ${\cal A}$ can be taken as:

\begin{equation}
g = \left( \matrix {U(x) & 0 \cr 0 & U^{\prime} } \right) \,\,\,,
\,\,\,\,\,\, U,U^{\prime } \in U(N)
\end{equation}
Then for the representation of $L$ in ${\cal H}$ we have:
\begin{eqnarray}
\pi(L) &=& g [\,\,D\,,\,g^* \,\,]
= \left ( \matrix{U \partial \kern -0.5em /  U^{\dag} & \gamma _5 U
(M\,U'^{\dag} \,-\,U^{\dag} \,M) \cr
\gamma _5 U'(M^{\dag }\,U^{\dag}\,-\,U'^{\dag}\,M^{\dag}) & 0 } \right ).
\end{eqnarray}

To obtain the effective chiral perturbation lagrangian, we first expand
the lagrangian in terms of derivatives of the field $g$ or powers of $L$.
\begin{equation}
{\cal L}_{eff}= {\cal L}^{(0)}+ {\cal L}^{(1)}+ {\cal L}^{(2)} + ... \,\,,
\end{equation}
where ${\cal L}^{(n)}$ are the set of terms which contain $n$ derivatives of
the field $g$.
${\cal L}^{(0)}$ should be set equal to  zero, because according to PCAC
all the interactions are momentum dependent\cite{L}. In this direction
${\cal L}^{(1)}$
through ${\cal L}^{(4)}$ will assume to have the following form:

\begin{eqnarray}
{\cal L}^{(1)} &=& Tr(K^{(1)}_1 L ) \\
{\cal L}^{(2)} &=& Tr(K^{(2)}_1 L^2 ) +[Tr(K^{(2)}_2 L )]^2  \\
{\cal L}^{(3)} &=& Tr(K^{(3)}_1 L^3 )+Tr(K^{(3)}_2 L )Tr(L^2)+
[Tr(K^{(3)}_3L)]^3  \\
{\cal L}^{(4)} &=& Tr(K^{(4)}_1 L^4 )+[Tr(K^{(4)}_2 L^2 )]^2 +
Tr(K^{(4)}_3 L ) Tr(L^3) \\
&+& Tr(K^{(4)}_4 L ) [Tr(L)]^3 +[Tr(K^{(4)}_5 L )]^4 +
[Tr(K^{(4)}_6 L )]^2 Tr(L^2) \,\,,
\end{eqnarray}
where $ K^{(i)}_j$ are some diagonal $2 \times 2$ matrices whose elements
$( k^{(i)}_j, k'^{(i)}_j)$  are the coupling
constants of the theory\footnote{One may assume $k^{(i)}_j$ is the coupling
constants on one layer and $k'^{(i)}_j$ is due to the other (infact the
disjoint point).}. It is interesting to note that the ${\cal L}^{(odd)}$s
vanish due to the vanishing of the trace of an odd number of Dirac marices.
Now if we take $ \chi =M U' M^{\dag}$ and directly calculate ${\cal L}^{(2)}$
and ${\cal L}^{(4)}$ and using the trace identity in ref\cite{GL},
we will obtain the effective lagrangian up to 4th order of momentum as follows:
\begin{eqnarray}
{\cal L}^{} & = & [-4k^{(2)}_{1}-8m^{2}(3k^{(4)}_{1}+ k'^{(4)}_{1})-
192 m^{2}k^{(4)}_{2}
(k^{(4)}_{2}+k'^{(4)}_{2})] \,\,Tr(\partial_{\mu}U \partial^{\mu}U^{\dag})\nn
\\
&+& [-4(k^{(2)}_{1}+k'^{(2)}_{1})-16m^{2}(k^{(4)}_{1}+ k'^{(4)}_{1})-192 m^{2}
(k_2^{(4)}+k'^{(4)}_2)^2] \,\,Tr(\chi U^{\dag}+U \chi^{\dag})\nn \\
&+&(-2 k^{(4)}_{1}+16(k^{(4)}_{2})^{2})\,\,
[Tr(\partial_{\mu}U \partial^{\mu}U^{\dag})]^{2}
-4k^{(4)}_{1}\,\, Tr(\partial^{\mu}U \partial^{\nu}U^{\dag})\,
Tr(\partial_{\mu}U \partial_{\nu}U^{\dag})\nn \\
&+&16 k^{(4)}_{1}
Tr(\partial_{\mu}U \partial^{\mu}U^{\dag}\partial_{\nu}U
\partial^{\nu}U^{\dag})
+32 k^{(4)}_{2}(k^{(4)}_{2}+k'^{(4)}_{2})\,\,Tr(\partial_{\mu}U
\partial^{\mu}U^{\dag})
\,\,Tr(\chi U^{\dag}+U \chi^{\dag})\nn \\
&+&4(3k^{(4)}_{1}+k'^{(4)}_{1})\,\,
Tr(\partial_{\mu}U \partial^{\mu}U^{\dag}(\chi U^{\dag}+U \chi^{\dag}))
+16( k^{(4)}_{2}+k'^{(4)}_{2})^{2}\,\,[Tr(\chi U^{\dag}+U
\chi^{\dag})\,]^{2}\nn\\
&+&4( k^{(4)}_{1}+k'^{(4)}_{1})\,\,Tr(\chi U^{\dag} \chi U^{\dag}+ \chi^{\dag}
U \chi^{\dag} U)
+8( k^{(4)}_{1}+ k'^{(4)}_{1})\,\,Tr(\chi \chi^{\dag})
\end{eqnarray}
which is the Gasser-Leutwyler Lagrangian\cite{GL} except for the external
fields . Here we concentrate on the symmetry breaking
term which are the $\chi $ dependent terms and ignore the external
fields\cite{GL,L}. By comparing (25) with the original
ChPL\cite{GL}, one can easily find the following relations:
\begin{eqnarray}
{1\over 4} F^{2}_{0} &=& -4k^{(2)}_{1}-8m^{2}(3k^{(4)}_{1}+
k'^{(4)}_{1})-192 m^{2}k^{(4)}_{2} (k^{(4)}_{2}+k'^{(4)}_{2}) \\
{1\over 4} F^{2}_{0} &=& -4(k^{(2)}_{1}+k'^{(2)}_{1})-16m^{2}(k^{(4)}_{1}+
 k'^{(4)}_{1})-192 m^{2} (k^{(4)}_{2}+k'^{(4)}_{2})^{2} \\
L_1 &=& -2k^{(4)}_{1}+16(k^{(4)}_{2})^{2} \\
L_2 &=& -4k^{(4)}_{1} \\
L_3 &=& 16 k^{(4)}_{1} \\
L_4 &=& 32 k^{(4)}_{2}(k^{(4)}_{2}+k'^{(4)}_{2}) \\
L_5 &=& 4(3k^{(4)}_{1}+k'^{(4)}_{1}) \\
L_6 &=& 16( k^{(4)}_{2}+k'^{(4)}_{2})^{2} \\
L_7 &=& 0 \\
L_8 &=& 4( k^{(4)}_{1}+k'^{(4)}_{1}) \\
H_2 &=& 8( k^{(4)}_{1}+ k'^{(4)}_{1}).
\end{eqnarray}
Now we can  see whether the relation (26)-(36) can at all fit the
experimental values given in Ref\cite{GL,EG} by a suitable choise of
$k$'s. A square fitting of $L$'s leads to the following numbers:
\begin{eqnarray}
k^{(4)}_1 &=& -0.24 \times 10^{-3} \cr
k'^{(4)}_1 &=& 0.78 \times 10^{-3} \cr
k^{(4)}_2 &\approx& k'^{(4)}_2 = i 1.84 \times 10^{-3} \;;\;i=\sqrt{-1}\,\,.
\end{eqnarray}
These values can be used as fit parameters in order to obtain small
deviation of $L_i$ parameters from their experimental values (in unit
$10^{-3}$)
\begin{eqnarray}
L_1 &=& 0.4 \;\;\;\;\;\; (0.7 \pm 0.3) \nonumber \\
L_2 &=& 1.0 \;\;\;\;\;\; (1.3 \pm 0.7) \nonumber \\
L_3 &=& -3.9\;\;\;\;\;\; (-4.4 \pm 2.5) \nonumber \\
L_4 &=& -0.2\;\;\;\;\;\; (-0.3 \pm 0.5) \nonumber \\
L_5 &=& 0.2 \;\;\;\;\;\; (1.4 \pm 0.5) \nonumber \\
L_6 &=& -0.2\;\;\;\;\;\; (-0.2 \pm 0.3) \nonumber \\
L_7 &=& 0   \;\;\;\;\;\; (-0.4 \pm 0.15) \nonumber \\
L_8 &=& 2.1 \;\;\;\;\;\; (0.9 \pm 0.3)\,\,,
\end{eqnarray}
where the values in parenthesis are the experimental values\cite{GL,EG}.
Also one can find some simple relations between $L_i$'s, for instance by
putting $k^{(4)}_2=k'^{(4)}_2$, we find:
\begin{eqnarray}
L_2 &=& {1 \over 2} (L_8 - L_5)\cr
L_3 &=& - 4 L_2\cr
L_4 &=& L_6 \cr
L_6 &=& 4 L_1 - 2 L_2 .
\end{eqnarray}

\section{Anomaly}
By now it seems that there is no possibility in our formalism to
obtain an anomaly cancellation term, (i.e. WZW term) naturally.
Besides, $L_7$ which is related to $U_A(1)$
anomaly\cite{A} is also zero. To solve the problem of WZW term, we should
change our geometrical space and take a 5-dimensional space.
Again we take the algebra ${\cal A}$ as (13), but this time ${\cal M}$ is
a 5-dimensional spin manifold. In fact we use the idea of Kaluza-Klein
for the continuous part of our geometrical space.
As is well known, there is no gradation matrix similar to $\gamma_5$ in the
5-dimensional space, so for Dirac operator we take it as follows:

\begin{equation}
D = \left( \matrix{ \partial \kern -0.5em /  \,\, \otimes \,\, { 1\kern - .4em
1}
& { 1\kern - .4em 1}'
\otimes M  \cr { 1\kern - .4em 1}' \otimes M^{\dag}  &
0 }
\right) \,\,,
\end{equation}
where $\partial \kern -0.5em / =\Gamma_{i} \partial^{i}$ and $\Gamma_{i}$
are the Dirac matrices for 5-dimensional space. By ${ 1\kern - .4em 1}'$, we
mean the unit matrix in internal spin space.
Now it is straightforward to calculate all the $ {\cal L}^{(i)}$s in the same
manner as we have done in the previous section, but up to 5th order of
momentum.
Again, we find that $ {\cal L}^{(1)}$ and $ {\cal L}^{(3)}$ will vanish, but
$ {\cal L}^{(5)}$ which is taken as:
\begin{equation}
{\cal L}^{(5)}=Tr(k^{(5)} L^{5})+(other \,\,terms \,\,in \,\,5th \,\,order
\,\,of \,\,momentum)
\end{equation}
will yield only one none vanishing term which is the WZW term,
\begin{equation}
{\cal L}^{(5)}= k^{(5)} \epsilon^{ijklm} Tr(U\partial_{i} U^{\dag}
\partial_{j} U \partial_{k} U^{\dag}\partial_{l} U \partial_{m} U^{\dag})
\,\,\,;\,i,j,k,l,m=0,1,2,3,5.
\end{equation}
This term is topological i.e. in action this term can be
integrated over the 4-dimensional space which
is the boundary of ${\cal M}$\cite{W}. Other terms ($ {\cal L}^{(2)}$ and $
{\cal L}^{(4)}$
) will have the same form as before except for the terms which contain
$\partial_5 U$. We will omit them because, we are in the low energy regime
and can  not observe the 5th dimension ( i.e. the same as in the  Kaluza-Klein
program).
By this strategy what will remain is the ChPL up to order four of
momentum and WZW term (except for the external field dependent terms).
Unfortunately the problem of $U_A(1)$ anomaly term will remain in this
approach.

\section *{Acknowledgements}

The authors wish to thank F. Ardalan, H. Arfaei and W. Nahm
for fruitful discussions.

\end{document}